\PassOptionsToPackage{table}{xcolor}
\documentclass[journal]{IEEEtran}

\usepackage{amsmath,amssymb}
\usepackage{textcomp}
\usepackage{graphicx}
\usepackage{booktabs}
\usepackage{array}
\usepackage{multirow}
\usepackage{makecell}
\usepackage{listings}
\usepackage[most]{tcolorbox}
\usepackage{xcolor}
\usepackage{subcaption}
\usepackage[numbers,sort&compress]{natbib}
\usepackage[hidelinks]{hyperref}

\def\BibTeX{{\rm B\kern-.05em{\sc i\kern-.025em b}\kern-.08emT\kern-.1667em\lower.7ex\hbox{E}\kern-.125emX}}
\newcolumntype{L}[1]{>{\raggedright\arraybackslash}p{#1}}
\newcolumntype{C}[1]{>{\centering\arraybackslash}p{#1}}
\lstdefinestyle{pythoncode}{
  language=Python,
  basicstyle=\ttfamily\footnotesize,
  keywordstyle=\color{blue!70!black},
  commentstyle=\color{green!40!black},
  stringstyle=\color{red!50!black},
  showstringspaces=false,
  columns=fullflexible,
  keepspaces=true,
  breaklines=true,
  breakatwhitespace=false,
  frame=single,
  framerule=0.3pt,
  rulecolor=\color{black!25},
  backgroundcolor=\color{black!3},
  xleftmargin=0.4em,
  aboveskip=0.5em,
  belowskip=0.5em
}

\lstdefinestyle{pascalcode}{
  language=Pascal,
  basicstyle=\ttfamily\footnotesize,
  keywordstyle=\color{blue!70!black},
  commentstyle=\color{green!40!black},
  stringstyle=\color{red!50!black},
  showstringspaces=false,
  columns=fullflexible,
  keepspaces=true,
  breaklines=true,
  breakatwhitespace=false,
  frame=single,
  framerule=0.3pt,
  rulecolor=\color{black!25},
  backgroundcolor=\color{black!3},
  xleftmargin=0.4em,
  aboveskip=0.5em,
  belowskip=0.5em,
  morekeywords={function,var,begin,end,then,else}
}

\usepackage{xspace}
\newcommand{\KLpair}{Knight\allowbreak--Leveson\xspace}

\begin{document}
\emergencystretch=2.5em\relax

\title{N-Version Programming with Coding Agents}

\author{Javier~Ron\textsuperscript{*},
        Benoit~Baudry\textsuperscript{\dag},
        and Martin~Monperrus\textsuperscript{*}\\
\textsuperscript{*}KTH Royal Institute of Technology\quad
\textsuperscript{\dag}Universit\'e de Montr\'eal\\
\texttt{javierro@kth.se}\quad
\texttt{benoit.baudry@umontreal.ca}\quad
\texttt{monperrus@kth.se}}

\maketitle

\begin{abstract}
This paper revisits the classical concept on N-version programming in the setting of contemporary AI coding agents. Revisiting the seminal \KLpair{} experiment, we study whether diversity across agent systems, models, and implementation languages creates diverse failure modes.
Using the \KLpair{}'s, Launch Interceptor Program Specification, we evaluate 48 agent-generated implementations on a shared oracle and a campaign of 1{,}000{,}000 randomized test inputs. The results show substantial common-mode failure, along the findings of \KLpair{}. Further analysis that many of those co-occuring failures can be traced to where is specification is particularly hard or ambiguous.
We also demonstrate that diversity from coding agents provides practical benefit: across majority voting three-version units, the mean failure count drops from 387.44 for single versions to 130.99 for triples, and 11{,}844 N-version units exhibit zero observed failures. Our original results is the strongest evidence to date that N-Version Programming with coding agents is a useful engineering strategy.
\end{abstract}

\section{Introduction}\label{sec:introduction}
N-version programming (NVP) promises reliability through diversity:
multiple independently produced implementations of the same specification are
executed in parallel, and a voting rule masks individual faults.
Its classical reliability argument, however, depends on a strong condition:
the versions must fail independently, or at least be diverse enough that
coincident failures remain rare.
That assumption was challenged in the human-programmer era:
the seminal \KLpair{} experiment showed that independently developed human implementations of the same specification still exhibited substantial common-mode failure~\cite{knight1986experimental}.

AI coding agents make this question relevant again~\cite{hassan2025agenticsoftwareengineeringfoundational, zheng2025can}.
Compared with recruiting independent human teams, it is now straightforward to
generate many implementations of the same task while varying three key components: the coding agent, the
underlying foundation model, and the target programming language.
With agentic coding, the central question of N-Version Programming remains unchanged: do the generated variants behave like
independent versions, or do they converge to the same latent defects?

This paper revisits the classical N-Version Programming literature in the modern agentic setting.
We design and perform a  reproduction of the original \KLpair{} experiment with AI coding
agents. We also revisit the broader questions about N-version software: whether the considered diversity mechanisms reduce fault
correlation, which shared fault families dominate, and whether redundancy helps even if fault independence fails. 

\begin{figure}[t]
  \centering
  \includegraphics[width=\linewidth]{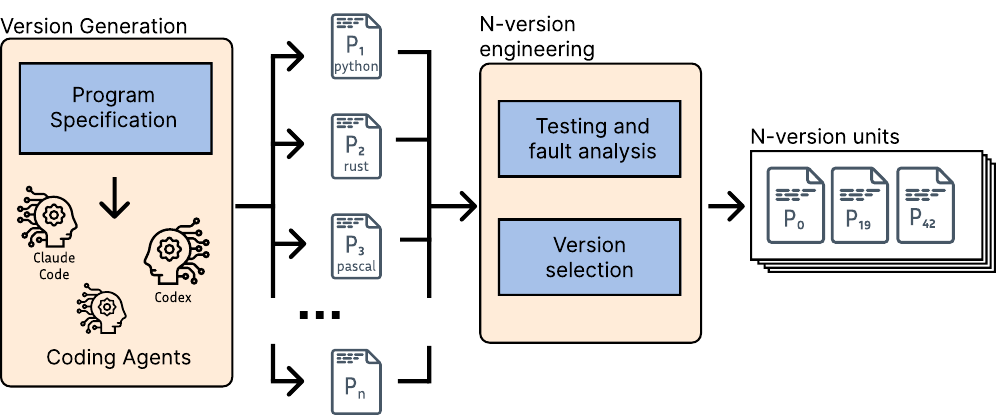}
  \caption{N-version units are generated by AI coding agents, they improve reliability over single programs on average.}
  \label{fig:pipeline}
\end{figure}

As specification, we take \KLpair{}'s one for the Launch
Interceptor Program (LIP), ane example defense software system.
We strictly follow the structure of the original experiment, only replacing human programmers with contemporary AI coding agents.
Across five agent systems, 23 models, and three target languages, our design
space comprises 69 \mbox{[harness, model, language]} triples.
We evaluate the resulting implementations with an oracle-based acceptance
screen and a shared 1{,}000{,}000-case campaign. We analyze the
observed failures through population-level coincident-failure statistics,
pairwise overlap across language and agent boundaries, function-level fault
localization, and exhaustive majority-vote three-version units.

Our results are clearcut. First, the idealized fault independence hypothesis fails: among the 48 admitted implementations in the campaign archive, the
experiment produces 429 coincident-failure cases where the random independence model
predicts only 115.36 ($z=29.20$), and strong pairwise clusters can be observed across
both language and agent boundaries.
The failures are not diffuse; they concentrate in a small number of recurring
bug families that reappear across ostensibly different implementations.
Second, despite fault correlation, redundancy does provide measurable benefit: when all
$\binom{48}{3}=17{,}296$ 3-Version units are evaluated, the mean failure count
drops from 387.44 for single versions to 130.99 for triples, and 11{,}844
triples exhibit zero observed failures.
All code and generated versions are available at \url{https://github.com/ASSERT-KTH/Knight-Leveson-Redux}

To sum up, the paper makes four contributions.
\begin{itemize}
    \item  First, it provides a systematic experimental framework for studying
fault independence and reliability in agent-generated software, faithful to \KLpair{}.
    \item  Second, it shows that modern coding agents can feasibly generate enough versions to cheaply perform N-Version programming  at scale.
    \item  Third, it measures and demonstrates failure overlap across \mbox{[harness, model, language]} diversity axes. Most of the errors can be traced back to weaknesses in the specification.
    \item  Fourth, it shows that majority voting N-version units do provide practical reliability gains, despite fault correlation.

\end{itemize}

\section{Background}\label{sec:background}

\begin{figure*}[t]
  \centering
  \includegraphics[width=\linewidth]{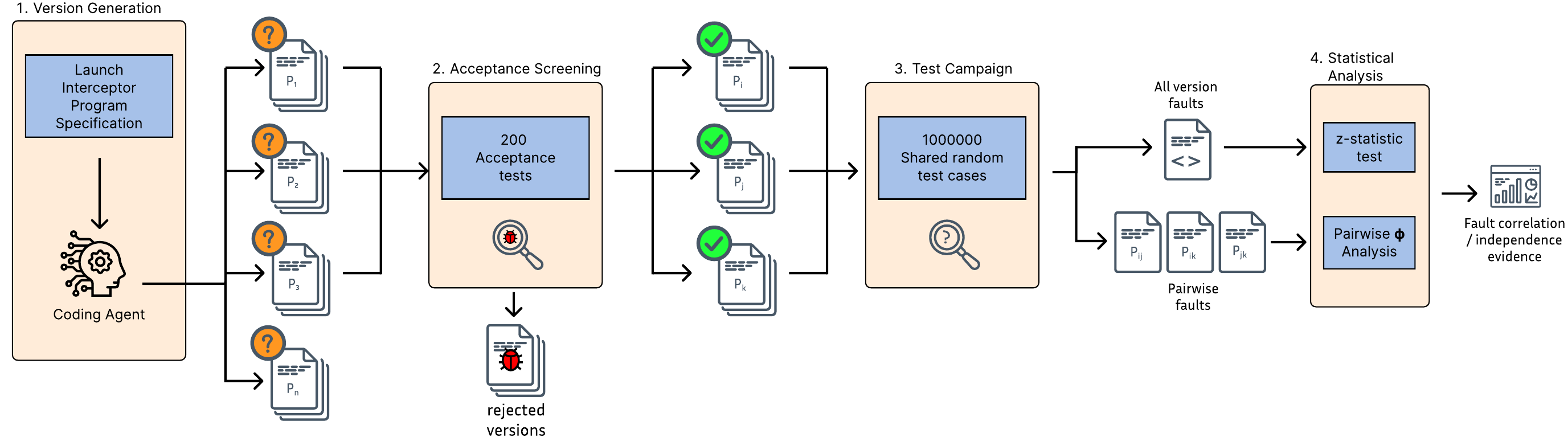}
  \caption{Experimental workflow for revisiting \KLpair{} with
           agentic coding: (1)~version generation by each agent against the LIP specification,
           (2)~oracle-based acceptance screening on 200 independently drawn
           random cases (all must pass),
           (3)~shared one-million test campaign across admitted versions, and (4)~statistical analysis
           (\KLpair{} $z$-statistics, pairwise correlation analysis). Versions that fail acceptance are
           excluded from the campaign and from all downstream analyses.}
  \label{fig:pipeline}
\end{figure*}

\subsection{N-Version Programming}

N-version programming was proposed by Chen and Avizienis~\cite{chen1978nversion}
as a software analog of hardware N-modular redundancy.
The design calls for $N$ independent teams to implement the same specification
from a common requirements document, then execute all versions in parallel on
each input and determine the output by majority vote or other consensus
mechanism.
Avizienis~\cite{avizienis1985nversion} formalized the reliability model:
provided that the failure events of distinct versions on any given input are
mutually independent and that individual failure probabilities are small,
the probability of majority failure decreases exponentially with $N$.

\subsection{N-Version and Fault Independence}
The idealized reliability benefit of NVP is entirely contingent on fault independence.
Eckhardt and Lee~\cite{eckhardt1985theoretical} provided a theoretical
treatment showing that independence cannot be assumed as a matter of
principle.
Because all versions are developed from the same specification, and because
specifications are finite and sometimes ambiguous, there exists a nonzero set
of inputs for which the specification is underspecified or misinterpreted in
a common way.
Programmers who share a training background, a programming language, or
exposure to the same reference materials will tend to make the same
misinterpretation, creating systematic coincident failure modes.
Littlewood and Miller~\cite{littlewood1989conceptual} extended this
analysis, arguing that the very process of translating a specification into
code creates correlations among any set of implementations derived from it.

\paragraph{The \KLpair{} Experiment (1986)}
The \KLpair{} experiment~\cite{knight1986experimental} was an
empirical test of fault independence in NVP.
Twenty-seven programmers from two universities independently implemented the
"Launch Interceptor Program" specification, working without mutual communication.
Their implementations were evaluated against a reference implementation on one million randomly drawn test cases.
Of the 24 versions with nonzero failure rates, failures were strongly
correlated: the observed count of simultaneous failures exceeded the
expectation under independence by a statistically significant margin.
Subsequent work by the same team~\cite{brilliant1990analysis} 
examined the coincidental faults, finding that a small number of
fault categories accounted for the majority of coincident events.
These faults were most prominently related to implementations of specific geometric computations required by the specification.
Hatton~\cite{hatton1997nversion} conducted a partial replication with a
different benchmark and similarly found that independence was not achieved.

\paragraph{Subsequent N-Version Work} Littlewood and Miller~\cite{littlewood1989conceptual} argued that
diversity mechanisms such as development methodology variation may improve reliability,
while Bishop~\cite{bishop1995design} identified ambiguities and omissions in
the specification as important sources of common-mode faults.
Hatton~\cite{hatton1997nversion} later showed that even without independence,
multi-version systems may still deliver useful reliability gains in practice.

\subsection{The Launch Interceptor Specification}
The specification for a `Launch Interceptor Program' (LIP) was defined by NASA  \cite{nasaLIP} and used in the original \KLpair{}
study.
The implementation task is to implement a \texttt{DECIDE} function that computes
a missile launch authorization decision, given $n$ planar
radar data points with Cartesian coordinates and a set of parameters representing incoming threats.
The implementation is not trivial because several functions involve non-trivial geometric computations that are
known fault attractors~\cite{brilliant1990analysis}.

The computation is specified with four stages:
(1)~the \emph{Conditions Met Vector} (\textsc{cmv}), a vector of 15 Boolean
Launch Interceptor Conditions (LICs), each encoding a geometric predicate on
subsets of the input points;
(2)~the \emph{Preliminary Unlocking Matrix} (\textsc{pum}), a $15{\times}15$
Boolean matrix derived from the \textsc{cmv} and a programmer-supplied Logical
Connector Matrix (\textsc{lcm}) whose entries are \texttt{ANDD},
\texttt{ORR}, or \texttt{NOTUSED};
(3)~the \emph{Final Unlocking Vector} (\textsc{fuv}), a 15-element
Boolean vector derived from \textsc{pum} column-wise conjunctions;
and (4)~the scalar \textsc{launch} decision, which is true iff all
\textsc{fuv} entries are true.

The Launch Interceptor Program 
is well suited to randomized testing: its input domain is (1) large, (2) the input types are simple (floating-point and enumeration parameters), and (3) random sampling covers the full condition space.

\subsection{Coding Agents as N-Version Generators}

Contemporary AI coding agents represent a qualitatively new kind of software
developer.
Chen et al.~\cite{chen2021codex} demonstrated that LLMs trained on code can
solve a substantial fraction of programming challenges; and modern coding agents can use tools, and perform iterative self-correction, and long-horizon
planning~\cite{jimenez2024swebench}.
Unlike earlier program synthesis systems, these agents are able to operate from general-purpose, natural language specifications.
Recent work such as Galapagos~\cite{ron2025galapagos} has begun to
exploit LLMs to construct functionally equivalent variants for N-version deployments.

Our present paper asks whether AI coding agents can generate diverse program
versions whose failures behave like the independent faults assumed by
classical N-version programming studies, or whether they reproduce the same
kinds of correlated failure modes those studies repeatedly found in practice.

\section{Experimental Methodology}\label{sec:methodology}
We design and perform an original experimental plan, that both replicates and extends the \KLpair{} experiment.

\subsection{Research Questions}

The core idea is to ask distinct coding agents to implement the same specification: LIP.
From there, we answer five research questions:

\medskip
\noindent\textbf{RQ1. (Generation Capabilities)} \textit{To what extent are AI coding agents able to implement the LIP specification?}

\vspace{7pt}

We first ask whether modern coding agents
can serve as generators of complete program versions from the seminal LIP specification. This inquiry is a necessary condition for collecting a sufficiently large and varied set of
working implementations for comparative analysis in the subsequent questions.

\medskip
\noindent\textbf{RQ2. (Fault Independance)} \textit{Are failures of AI-generated implementations mutually independent, per idealized random fault model?}

\vspace{7pt}

This is the central question of the original \KLpair{} study, translated to
the agentic world. We test whether simultaneous failures
across generated implementations do not happen (consistent with independence), or whether
joint failures occur substantially more often than that model predicts (consistent with independence).

Our experimental design follows the
\KLpair{} methodology as closely as the agentic setting permits. Table~\ref{tab:kl_comparison} summarizes the relationship
between the original study and the present one.

\medskip
 \noindent\textbf{RQ3. (Diversity Dimensions)} \textit{To what extent does diversity across programming languages, coding agents, and AI models impact fault independence?}

\vspace{7pt}

Littlewood and Miller~\cite{littlewood1989conceptual}
 argued that diverse software generation mechanisms, such as development methodology, 
may improve the dependability of multi-version systems. We therefore
ask whether varying the language, agent, or underlying model
decorrelates failures, or whether shared failure modes remain dominant across
these dimensions.

\medskip
\noindent\textbf{RQ4. (Fault Explanation)} \textit{What are the principal sources of faults in AI-generated implementations?}

\vspace{7pt}

Beyond quantitatively measuring correlation, we qualitatively investigate the origin of failures. We study whether faults in agent-generated
programs can be attributed to a small number of challenging or
ambiguous parts of the specification.

\medskip
\noindent\textbf{RQ5. (N-version Reliability)} \textit{To what extent do N-version units generated by AI agents provide practical reliability benefits?}

\vspace{7pt}

Failure correlation falsifies the ideal case for N-version programming, but does
not by itself imply that N-version is useless. Following the perspective of
Hatton's study~\cite{hatton1997nversion}, we measure whether combining multiple
AI-generated versions yields worthwhile error reduction.

\subsection{Methodology Overview}

At a high level, we answer the research questions by reproducing the \KLpair{}
experimental structure faithfully.
Table~\ref{tab:kl_comparison} summarizes which elements are carried over
directly and the extent of which the experimental design has been adapted for AI coding agents.
As summarized in Table~\ref{tab:kl_comparison}, we preserve the original
acceptance filter, campaign size, failure definition, and primary
\KLpair{} hypothesis test.

\begin{table*}[t]
  \renewcommand{\arraystretch}{1.25}
  \centering
  \begin{tabular}{L{3.2cm} L{6.2cm} L{6.2cm}}
    \toprule
    \textbf{Dimension} &
    \textbf{\KLpair{} (1986)} &
    \textbf{This Study} \\
    \midrule
    Subjects &
    27 human programmers (graduate and undergraduate students at 2 universities) &
    69 coding agents over 5 harnesses (Cursor, Claude Code, OpenAI Codex, Gemini,
    OpenCode)\\[2pt]

    Implementation Language  &
    Pascal &
    Pascal, \emph{Python, Rust} \\[2pt]
    
    Specification &
    Original LIP specification document + 15 input/output examples + \texttt{realcompare} function &
    Faithful; Original LIP specification document + 15 input/output examples + \texttt{realcompare} function + \emph{a prompt of the task to perform} \\[2pt]

    Admitted version count &
    27 &
    48 \\[2pt]

    Test-campaign size &
    1{,}000{,}000 random cases &
    Faithful: 1{,}000{,}000 random cases \\[2pt]

    Acceptance filter &
    200-case pre-screen; all must pass &
    Faithful: 200-case pre-screen; all must pass \\[2pt]

    Oracle &
    Pascal reference implementation &
    Python reference implementation validated by
    82 unit tests covering the full spec. \\[2pt]

    Failure definition &
    Any of 241 output bits differs from oracle &
    Faithful: any of 241 output bits differs from oracle \\[2pt]

	    Primary statistic &
	    \KLpair{} $z$-statistic (Eq.~\ref{eq:z}) &
	    Faithful; \KLpair{} $z$-statistic, \emph{supplemented by pairwise correlation analysis, cross-language and cross-agent stratification, root cause analysis, and majority-vote unit analysis} \\[2pt]

    Independence enforcement &
    Different universities, no communication between students &
    Distinct agent harnesses and vendors; distinct underlying model
    lineages; distinct target languages \\[2pt]

    Primary confounders &
    Shared
    specification document, shared university curriculum  &
    Faithful; shared
    specification document, \emph{shared LLM training corpora, overlapping model lineages} \\
    \bottomrule
  \end{tabular}
  \caption{Relationship to \KLpair{} (1986): faithful methodological elements, adaptations and \emph{novelty (in italics)}.}
  \label{tab:kl_comparison}
\end{table*}

\noindent\textbf{RQ1. }First, we adapt the source of diversity from
human programmers to \mbox{[harness, model, language]} configurations.
We generate candidate versions from a controlled set of
\mbox{[harness, model, language]} configurations and apply the original \KLpair{}-style
acceptance screen; the number of admitted versions is the feasibility evidence of agentic N-Version programming. 

\noindent\textbf{RQ2. }Second, we run every admitted version on the same million-input 
campaign and record a binary failure vector for each implementation; these
vectors are the core data for the independence test. 

\noindent\textbf{RQ3. }Third, we stratify the pairwise co-failure data by
language pair and by same-agent versus cross-agent comparisons to test whether
these diversity dimensions reduce overlap in failure behavior.

\noindent\textbf{RQ4. }Fourth, for failed executions we retain
the triggering inputs and output mismatches, then aggregate them at LIC level to determine the major failure
sources and trace back the problem to the specification. 

\noindent\textbf{RQ5. }Finally, we reuse the same
failure vectors to simulate majority-vote N-version units, and quantitatively measure the reliability improvements brought by agentic N-version.

\subsection{Version Generation with LLMs}

Five AI coding agent systems serve as the ``programmers'' in our study:
Cursor~\cite{cursor2024}, Claude Code~\cite{anthropic2024claude}, OpenAI
Codex~\cite{chen2021codex}, Gemini~\cite{gemini2024}, and
OpenCode~\cite{opencode2024}.
Each agent is configured with a list of underlying models spanning, where
applicable, multiple vendors and generations as described in Table~\ref{tab:agent_models}.
We configure Cursor with the Composer models;
Claude Code with Anthropic's Haiku, Sonnet, and Opus variants;
Codex with multiple GPT-5.x revisions;
Gemini with its 2.5 and 3.x preview variants;
and OpenCode with Qwen and Gemma models.

The functional specification given to every agent is the original \KLpair{} specification document,
preserved verbatim as the authoritative source of
truth for all conditions, LICs, and \texttt{realcompare} semantics.
Agents are additionally given: a file with 15 input/output examples, and a reference \texttt{realcompare} implementation.
The agents also receive a short directive that describes the provided information, input and output formats, and the expected deliverable 
(\href{https://github.com/ASSERT-KTH/Knight-Leveson-Redux/blob/fa95aebe1dd888b69ca594f059433d08504bbde6/harnesses/python_harness.py#L14}{Python},
\href{https://github.com/ASSERT-KTH/Knight-Leveson-Redux/blob/fa95aebe1dd888b69ca594f059433d08504bbde6/harnesses/rust_harness.py#L1}{Rust},
or \href{https://github.com/ASSERT-KTH/Knight-Leveson-Redux/blob/fa95aebe1dd888b69ca594f059433d08504bbde6/harnesses/pascal_harness.py#L1}{Pascal}).
No algorithm is suggested in the prompt, no code skeletons are provided.

\begin{table*}[t]
  \renewcommand{\arraystretch}{1.15}
  \centering
  \small
  \rowcolors{2}{white}{gray!10}
  \begin{tabular}{ll}
    \toprule
    \textbf{Harness} & \textbf{Models} \\
    \midrule
    Cursor &
    \texttt{composer-2.5}, \texttt{composer-2} \\
    Claude Code &
    \makecell[l]{\texttt{anthropic/claude-opus-4.6}, \texttt{anthropic/claude-opus-4.5},
    \\ \texttt{anthropic/claude-sonnet-4.6}, \texttt{anthropic/claude-sonnet-4.5},
    \\ \texttt{anthropic/claude-haiku-4.5}} \\
    OpenAI Codex &
    \makecell[l]{\texttt{gpt-5.4}, \texttt{gpt-5.4-mini}, \texttt{gpt-5.3-codex},
    \\ \texttt{gpt-5.2-codex}, \texttt{gpt-5.2}} \\
    Gemini &
    \makecell[l]{\texttt{gemini-3.1-pro-preview}, \texttt{gemini-3-flash-preview},
    \\ \texttt{gemini-2.5-pro}, \texttt{gemini-2.5-flash},
    \texttt{gemini-2.5-flash-lite}} \\
    OpenCode &
    \makecell[l]{\texttt{qwen/qwen3.6-plus}, \texttt{qwen/qwen3.5-flash-02-23},
    \texttt{qwen/qwen3.5-plus-02-15}, \\ \texttt{qwen/qwen3.5-397b-a17b},
    \texttt{google/gemma-4-26b-a4b-it}, \texttt{google/gemma-4-31b-it}} \\
    \bottomrule
  \end{tabular}
  \caption{Coding harnesses and underlying models used for version generation. A coding agent is a combination of a harness and a model.}
  \label{tab:agent_models}
\end{table*}

We pair each \mbox{[harness, model]} tuple with each target programming language,
producing three corresponding \mbox{[harness, model, language]} triples.
Each triple produces one candidate \texttt{DECIDE} program per invocation.
The resulting pool of versions therefore consists of \textbf{69} unique \mbox{[harness, model, language]} runs.
The exact agent system, underlying model identifier, and target language
are recorded in the per-version metadata for reproducibility.
This metadata is later used to separate failures by agent and language
when answering RQ3.

\subsection{Oracle and Acceptance Testing}
This part of the methodology defines the reference implementation and the
admission filter that determines which generated versions are eligible for the
main campaign.

We develop a reference implementation of \texttt{DECIDE} in Python,
validated by an automated test suite of 82 unit tests covering all known boundary conditions for the 15 LICs.
This reference implementation serves as the oracle for differential
testing: on every campaign input, each admitted version and the oracle
implementation are both evaluated, and the version is recorded as failing
on that case if any of its 241 output bits differs from the oracle's.
This matches the failure definition used by \KLpair{}.

Before entering the main test campaign (see~\ref{sec:main-campaign}), each
generated version undergoes an acceptance screening, using the terminology of the original \KLpair{} protocol.

Acceptance testing works as follows.
First, we check Pascal and Rust implementations for compilation errors.
Then, we use two hundred test cases and evaluate each version against the oracle.
Compilation and the small set of tests are used to filter out clearly non-functional implementations.
As in \KLpair{}, a version is admitted only if it passes all 200 tests.
The admitted-version count is the first empirical outcome of the study: it
measures whether AI coding agents can produce  complete-enough implementations to support the replicated N-version experiment in RQ2.

\subsection{Main Test Campaign}
\label{sec:main-campaign}

At this stage, versions that crash immediately or fail any acceptance test have been filtered out.
Therefore, the main campaign will only analyze implementation-level disagreements among the admitted versions.
We draw uniformly $T = 1{,}000{,}000$ test cases at random from the input domain using a fixed seed to ensure reproducibility.
All admitted versions are evaluated on exactly the same set $T$ against the oracle.
For all test runs, we record metadata: coding agent, model, language, and a
binary pass/fail outcome.
The resulting failure data is the common measurement for the rest of the paper:
(1) aggregate coincident failures are used for the statistical analysis in RQ2;
(2) stratified pairwise overlaps are used for the cross-language and
cross-agent analysis in RQ3; and (3) majority-vote simulations are used for RQ5.

For \textit{failed} test runs, we also record the input that caused the failure,
as well as the specific CMV, PUM, FUV, or LAUNCH oracle mismatches.
These fault records are later aggregated at LIC level and linked back to
representative implementations and triggering inputs for RQ4.

\subsection{Failure Correlation Statistical Analysis}
We implement an exact replicate of the \KLpair{} statistical framework
~\cite{knight1986experimental} to answer RQ2.
Let $N$ be the number of admitted versions, $T$ the number of campaign test
cases, $f_i$ the failure count for version $i$, and
$p_i = f_i / T$ the empirical failure rate.
Let $K$ denote the number of test cases on which \emph{two or more} versions
fail simultaneously.
$K$ is the aggregate measure of failure overlap: it counts how
often the test campaign encounters an input on which independent implementations
break at the same time.

We use the $z$-statistic to test the observed distribution of failures against 
an approximate normal distribution of failures.
Specifically, under the null hypothesis H$_0$ that failures are mutually independent
Bernoulli events:
\begin{align}
  P_0 &= \prod_{i=1}^{N}(1 - p_i) \label{eq:p0}\\
  P_1 &= \sum_{i=1}^{N}\Bigl[p_i \prod_{j \neq i}(1 - p_j)\Bigr] \label{eq:p1}\\
  P_m &= 1 - P_0 - P_1 \label{eq:pm}
\end{align}
where $P_0$, $P_1$, $P_m$ are the probabilities that exactly zero, one, or
two or more versions fail on a given test case.
In particular, $P_m$ is the quantity of interest under H$_0$: it is the
independence-based prediction for the chance that a test case triggers a
coincident failure event.
Under H$_0$, $K$ is approximately normally distributed with
mean $\mu = T P_m$ and standard deviation $\sigma = \sqrt{T P_m (1-P_m)}$,
giving the \KLpair{} $z$-statistic:
\begin{equation}
  z = \frac{K - \mu}{\sigma}. \label{eq:z}
\end{equation}
Because $K$ is taken to be approximately normal under H$_0$, tail probabilities and two-sided
$p$-values for the observed $z$ follow from the standard normal distribution.
We reject H$_0$ at the 99\% confidence level if $|z| > 2.576$,
matching the \KLpair{} threshold.
The alternative hypothesis H$_1$ is that failures are \emph{positively}
correlated, i.e. coincident failures exceed the independence prediction.
A large positive $z$ therefore means that the observed overlap in failures is
many standard deviations above what this independence model  would predict.

In addition to the pooled and per-language \KLpair{} counts, we compute a
global all-pairs similarity view over the admitted population.
For every pair of admitted versions, we compute the Pearson $\phi$
correlation between their binary failure vectors over the million-case
campaign.
Here, $\phi$ measures co-failure similarity:
$\phi = 1$ means two versions fail on exactly the same campaign inputs,
$\phi \approx 0$ means little overlap in their failure patterns, and negative
values indicate anticorrelation.
We use $\phi$ as the main descriptive overlap measure between pairs.
This complements the aggregate \KLpair{} statistic by showing whether coincident
failures are concentrated in clusters of versions or dispersed across the
population.

\subsection{Cross-Language and Cross-Agent Failure Analysis}
\label{sec:methodology-rq3}
RQ3 asks whether implementation language and coding agent behave as meaningful
axes of diversity, rather than as wrappers around the same failure
mode.
We answer that question by reusing the full pairwise failure data from RQ2 and
stratifying it along the diversity dimensions directly available in the
experiment.

To study language diversity, we partition the full pairwise set into
cross-language pairs and inspect their $\phi$ distribution.
This asks whether changing implementation language tends to decorrelate failure
behavior in the observed population.

Third, to study agent diversity, we partition the same pairwise set into
same-agent and cross-agent subsets and compare their $\phi$ distributions.
This asks whether crossing an agent boundary decorrelates failures, or whether
similar failure profiles remain common even across different tools.

\subsection{Root Cause Analysis}
We identify the sources of correlated faults by manually analyzing failures patterns.
Because the 15 LIC predicates form the main functional decomposition inside
\texttt{DECIDE}, they provide a natural first unit for fault localization.
First, we count how many distinct \mbox{[harness, model, language]} triples fail on
each LIC, and we stratify those LIC-level counts by target language and by
coding agent.
Then, we trace those failures back to implementation choices in the
generated source code, and compare against the oracle and the
specification.

\subsection{N-Version Unit Analysis}

We construct synthetic N-version units from the admitted
versions and evaluate them on the same $T=1{,}000{,}000$ campaign inputs. A unit
of size $n$ fails on an input when at least $\lceil n/2\rceil$ of its members
fail on that input.

For the core RQ5 analysis, we enumerate all possible three-version units that
can be formed from the admitted pool and compute their majority-vote failure
statistics over the campaign.
This gives an exhaustive picture of how often N-version voting helps improve reliability in the observed population.

\section{Experimental Results}\label{sec:results}

\subsection{RQ1: Acceptance Success}

\begin{figure}[t]
  \centering
  \includegraphics[width=\linewidth]{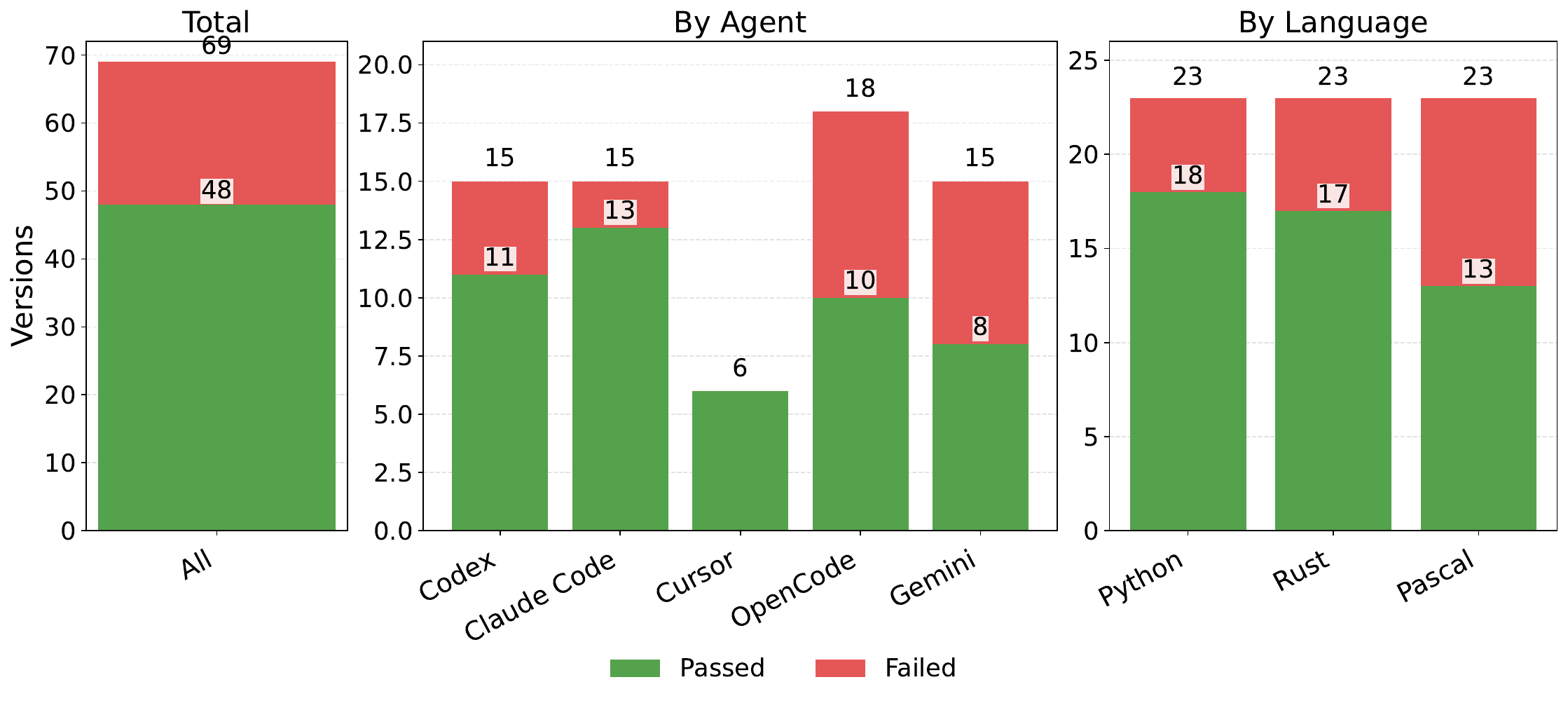}
  \caption{RQ1. Acceptance testing summary.
           Left: the total configured population of 69 triples.
           Center: counts by agent system.
           Right: counts by target language.
           In all panels, green denotes admitted versions and red denotes
           excluded versions; the stacked height is the total number of
           configured triples in that category.}
  \label{fig:admission}
\end{figure}

Fig.~\ref{fig:admission} summarizes the outcome of acceptance testing.
Across the five agent systems and three target languages, 69 configured
\mbox{[harness, model, language]} triples were generated and 48/69 passed all 200
acceptance cases, yielding an admission rate of 70\%.
The remaining 21 configured triples were excluded before the campaign.
The admission rate varies markedly by agent (center panel): the \textbf{Cursor programs are admitted for
all 6 of 6 attempts} (100\%) across all three languages, compared with
13 of 15 (87\%) for Claude Code, 11 of 15 (73\%) for Codex,
8 of 15 (53\%) for Gemini, and 10 of 18 (56\%) for OpenCode.
Similar variation was observed per language (right panel): \textbf{Python is the most
successful target language, with 18/23 admitted attempts} (78\%), while Rust has
17 of 23 (74\%) and Pascal 13 of 23 (57\%).

\emph{Agent Failure Modes}. 
Among the 21 excluded triples, 12 failed to produce the required artifact.
Among the remaining excluded versions, 5 crashed immediately when executed, and 4 returned wrong output on at least one acceptance case.
Consistent with the original \KLpair{} acceptance protocol, all such versions are excluded from the campaign and from all subsequent analyses.

\begin{tcolorbox}[
  enhanced,
  colback=black!6,
  colframe=black!35,
  boxrule=0.4pt,
  arc=2pt,
  left=6pt,
  right=6pt,
  top=6pt,
  bottom=6pt
]
\noindent\textbf{Answer to RQ1.}
AI coding agents can correctly implement the specification of the
 \KLpair{} study. We obtained 48 functional implementations which  pass the 200-case acceptance screen. The pool of admitted implementations is large enough to compute the \KLpair{}
independence test and to support additional analyses by language, harness,
model, and fault source. 
\end{tcolorbox}

\subsection{RQ2: Coincident Failure Statistics}
\label{sec:kl_stats}

\begin{figure}[t]
  \centering
\includegraphics[width=\linewidth]{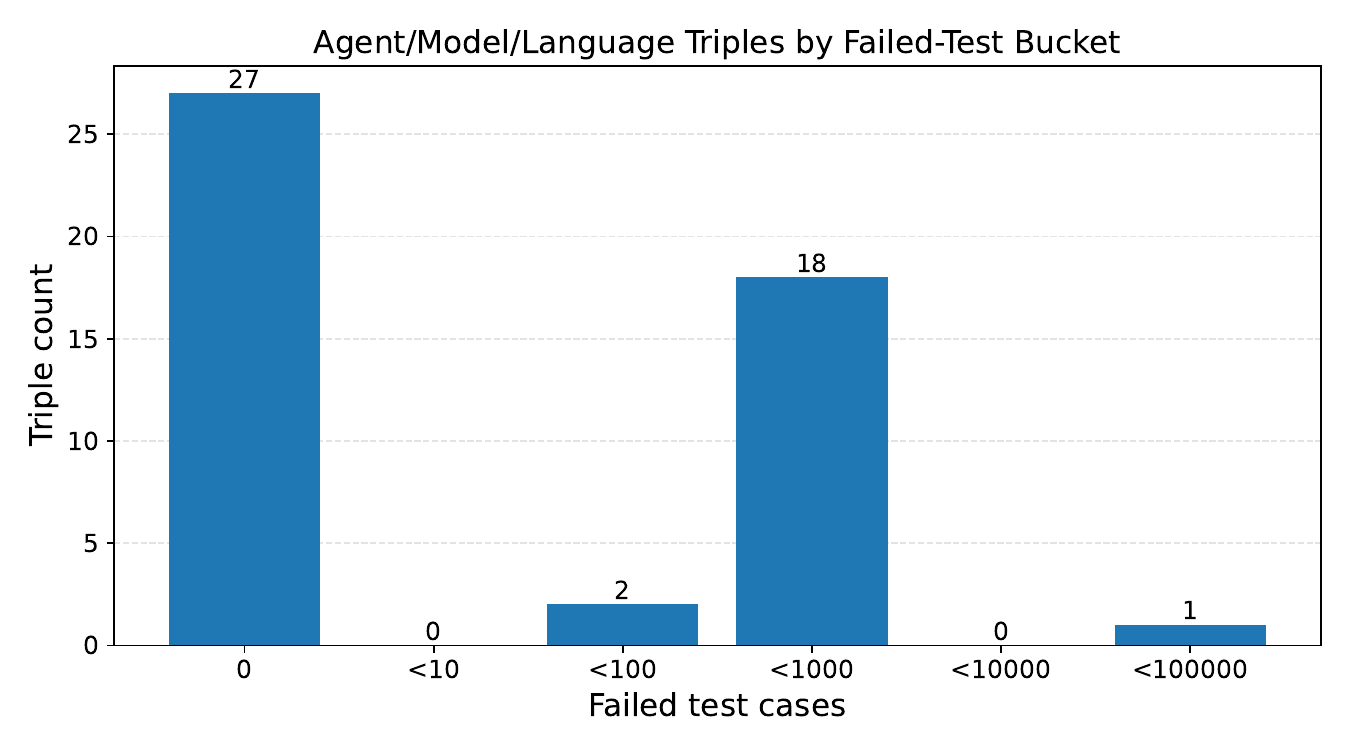}
  \caption{Failure-count distribution across the 48 admitted
           \mbox{[harness, model, language]} according to the reference implementation, grouped into buckets over the million test case campaign.
           The distribution is strongly right-tailed, with many near-perfect
           versions and a much smaller set of high-failure outliers.}
  \label{fig:failure_rates}
\end{figure}

The admitted versions span a wide range of failure counts in the main campaign: 27 are failure-free, while the worst version fails 10{,}469 of the $10^6$ inputs.
Fig.~\ref{fig:failure_rates} groups the 48 admitted
\mbox{[harness, model, language]} triples into coarse failure-count buckets.
Most versions cluster at exactly zero failures; 2 more fall between 10 and 99 failures,
18 fall between $10^2$ and $10^3$ cases, none fall between $10^3$ and $10^4$,
and 1 exceeds $10^4$ failures.
The right tail shows that the agent population contains both a large near-perfect core and a smaller set of high-failure outliers.

Table~\ref{tab:kl_stats} presents the \KLpair{} statistics for the whole
48-version campaign.
The \KLpair{} test asks whether the observed number of campaign inputs
on which at least two versions fail simultaneously is compatible with the
independence model derived from the individual version failure rates.
If the observed count $K$ is much larger than the independence expectation
$\mu$, the resulting large positive $z$ indicates common-mode failure rather
than accidental overlap.

Pooled across all languages, the independence model predicts
$\mu = 115.36$ coincident-failure cases over the $T = 10^6$ campaign;
the observed count is $K = 429$, an excess of $K/\mu \approx 3.7\times$
over the independence prediction, yielding $z = 29.20$ with $p \approx 1.765 \times 10^{-187}$.
This infinitesimal p-value decisively rejects the independence hypothesis.

Next, we do a per-language analysis. Every per-language analysis rejects H$_0$ with over 99.99\% confidence, with $K/\mu$ ranging from $17.5\times$
(Python) to $155.1\times$ (Pascal) and $z$ from 80.7 to 253.3.
The systematic rejection rules out the explanation that the pooled rejection is being driven by only one implementation language.

The fact that Z is lower at the whole population level shows that the pooled population is more heterogeneous than the within-language slices, reducing the relative concentration of coincident failures in the aggregate.

We next turn from coincident-failure counts to pairwise failure similarity.
This global all-pairs view shows how the 48 admitted versions relate to one another as a population and whether the same dependence appears beyond the pooled \KLpair{} statistic.
For every pair of admitted versions, we compute the Pearson $\phi$
correlation between their binary failure vectors over the main test campaign.
In this setting, $\phi$ measures co-failure similarity:
$\phi = 1$ means two versions fail on exactly the same inputs,
$\phi \approx 0$ means little overlap in their failure patterns,
and negative values would mean their failures are anticorrelated.

\begin{figure*}[t]
  \centering
  \includegraphics[width=\textwidth]{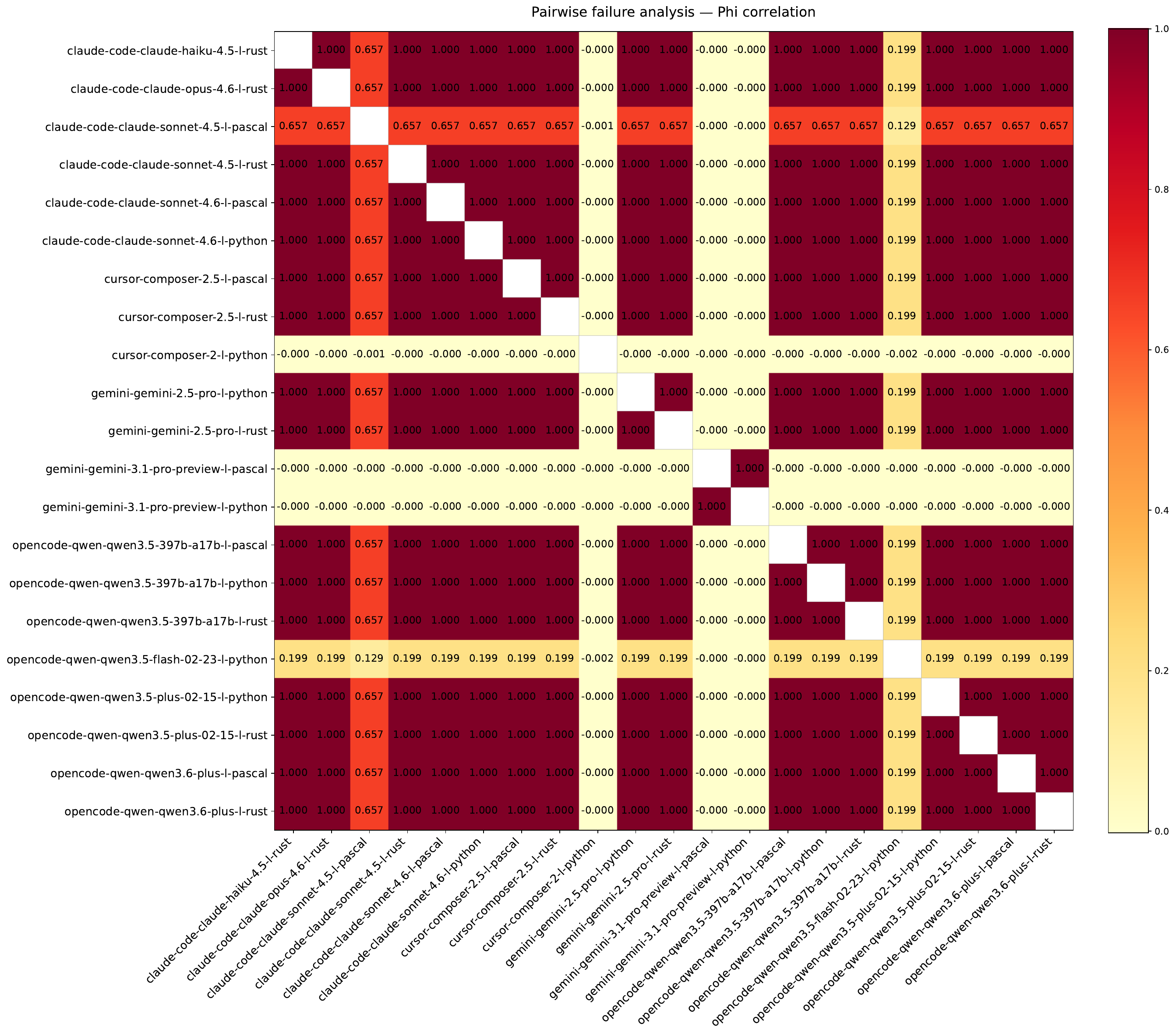}
  \caption{Filtered pairwise co-failure heatmap for the 21 versions
           with at least one observed campaign failure.
           Each cell shows the Pearson $\phi$ correlation between the
           binary failure vectors of two versions; rows and columns are ordered
           by agent system and then by language.
           Dark blocks correspond to
           clusters of versions that fail on exactly
           the same campaign inputs. The dark zones cross agent and language
           boundaries.}
  \label{fig:heatmap}
\end{figure*}

Fig.~\ref{fig:heatmap} shows a filtered heatmap of the pairwise $\phi$ matrix,
restricted to versions with at least one observed campaign failure.
Large dark regions indicate families of versions that fail on the same inputs, and those regions cross both agent and language boundaries
rather than aligning neatly with a single tool or target language.
Again, the immediate implication is that nominal diversity in agent, model, or
language does not automatically buy behavioral diversity.
Pale rows and columns in
the heatmap mark failing implementations whose failures do not overlap with the
main co-failure clusters.

\begin{tcolorbox}[
  enhanced,
  colback=black!6,
  colframe=black!35,
  boxrule=0.4pt,
  arc=2pt,
  left=6pt,
  right=6pt,
  top=6pt,
  bottom=6pt
]
\noindent\textbf{Answer to RQ2.}
The generated AI-agent implementations do \emph{not} fail independently, according to a uniform random model.
At the pooled level, the experiment produces 429 coincident-failure cases where the independence model predicts only 115.36; the null hypothesis is rejected with $p \approx 1.765 \times 10^{-187}$.
\end{tcolorbox}


\begin{table}
  \renewcommand{\arraystretch}{1.0}
  \centering
  \small
  \begin{tabular}{@{}lrrrrr@{}}
    \toprule
    \textbf{Slice} & $N$ & $\mu$ & $K$ & $z$ & $p$-value \\
    \midrule
    All languages  & 48 & 115.36 & 429 & 29.20 & $1.765 \times 10^{-187}$ \\
    Python         & 18 &  23.97 & 419 &  80.69 & $1.701 \times 10^{-1416}$ \\
    Rust           & 17 &   4.91 & 419 & 186.93 & $1.693 \times 10^{-7590}$ \\
    Pascal         & 13 &   2.70 & 419 & 253.30 & $3.327 \times 10^{-13935}$ \\
    \bottomrule
  \end{tabular}
  \caption{RQ2. \KLpair{} correlated failure statistics.
           $N$ is the number of versions;
           $\mu = T \cdot P_m$ is the expected simultaneous-failure  count according to the Bernoulli model;
           $K$ is the observed count in the respective population;
           $z$ is the $z$-statistic
           (see Eq.~\ref{eq:z}); and $p$ is the corresponding two-tailed p-value.
           All slices decisively reject H$_0$: the theoretical random failure model and the actual failure mode do not match.}
  \label{tab:kl_stats}
\end{table}

\subsection{RQ3: Cross-Language and Cross-Agent Failure Analysis}
\label{sec:results-rq3}
RQ3 studies whether language and agent variance produce diversity in the observed failure behavior.
We answer it by splitting the pairwise co-failure analysis along the diversity dimensions directly available in the experiment: implementation language and coding agent.

We first refine the global picture from RQ2 by looking specifically at cross-language pairs.
Here, every admitted version written in one language is paired with every
admitted version written in a different language, regardless of agent or model,
and we compute the same co-failure statistics as in the global analysis.
This yields 761 cross-language pairs in total; for 615 of them,
$\phi$ is undefined because at least one of the two versions is failure-free,
leaving 146 pairs in the $\phi$ distribution analysis.

\begin{figure}[t]
  \centering
  \includegraphics[width=\linewidth]{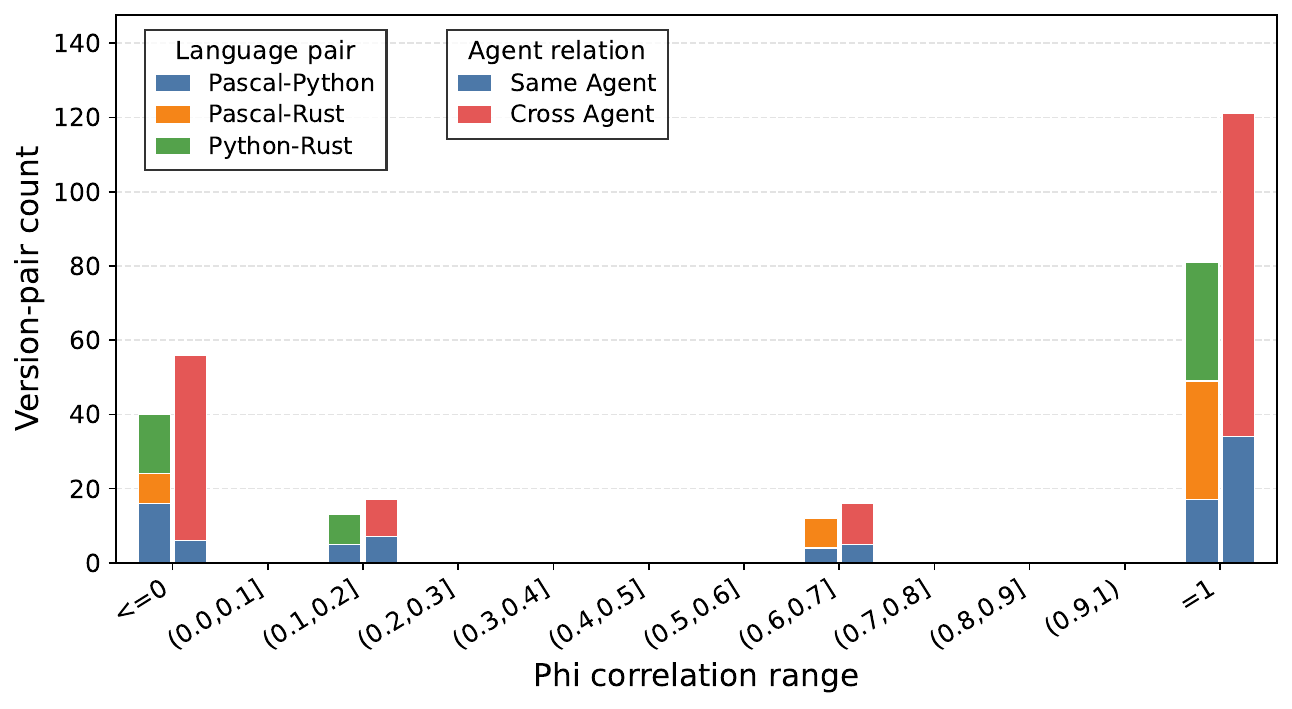}
  \caption{RQ3: Bucketed distributions of pairwise $\phi$ correlations under two diversity axes.
           For each bucket, the left stacked bar counts cross-language pairs by language pair, while the right stacked bar counts all version pairs by same-agent versus cross-agent relation.
           The exact-match bucket $\phi = 1$ is separated from the near-perfect bucket $(0.9,1)$, showing that the high-correlation mass is dominated by correlated failures.}
  \label{fig:phi_diversity_splits}
\end{figure}

Figure~\ref{fig:phi_diversity_splits} summarizes the resulting pairwise $\phi$ distributions under both diversity splits.
For cross-language pairs, 146 pairs have defined $\phi$; 81 land exactly in the $\phi = 1$ bucket, indicating perfect co-failure agreement.
The remaining defined cross-language pairs are fewer: 40 are non-positive, 13 fall in $(0.1, 0.2]$, and 12 fall in $(0.6, 0.7]$.
The language-pair stack shows that perfect co-failures are distributed across all three language pairings: Pascal--Python contributes 17 pairs, Pascal--Rust contributes 32, and Python--Rust contributes 32.

The same figure also shows the corresponding same-agent versus cross-agent split.
The observed pattern is similar in both groups: among the 221 same-agent pairs, 52 have defined $\phi$, with 34 at exact $\phi = 1$ and 6 non-positive; among the 907 cross-agent pairs, 158 have defined $\phi$, with 87 at exact $\phi = 1$ and 50 non-positive.
Crossing an agent boundary therefore does not eliminate highly correlated failure profiles: exact co-failure clusters remain common even between different tools, although the population also contains many genuinely distinct cross-agent pairs.

\begin{tcolorbox}[
  enhanced,
  colback=black!6,
  colframe=black!35,
  boxrule=0.4pt,
  arc=2pt,
  left=6pt,
  right=6pt,
  top=6pt,
  bottom=6pt
]
\noindent\textbf{Answer to RQ3.}
Cross-language and cross-agent program generation do not provide enough diversity to make failure correlation disappear.
In Littlewood and Miller's terms~\cite{littlewood1989conceptual}, varying the \mbox{[harness, model, language]} tuple does not equate to achieving fundamentally different methodologies that generate diverse program distributions.
\end{tcolorbox}

\subsection{RQ4: Root Cause Analysis}
\label{sec:results-rq4}
Next, we take a deeper look at the failure modes on the 1M random test cases.

\begin{figure}[t]
  \centering
  \includegraphics[width=\linewidth]{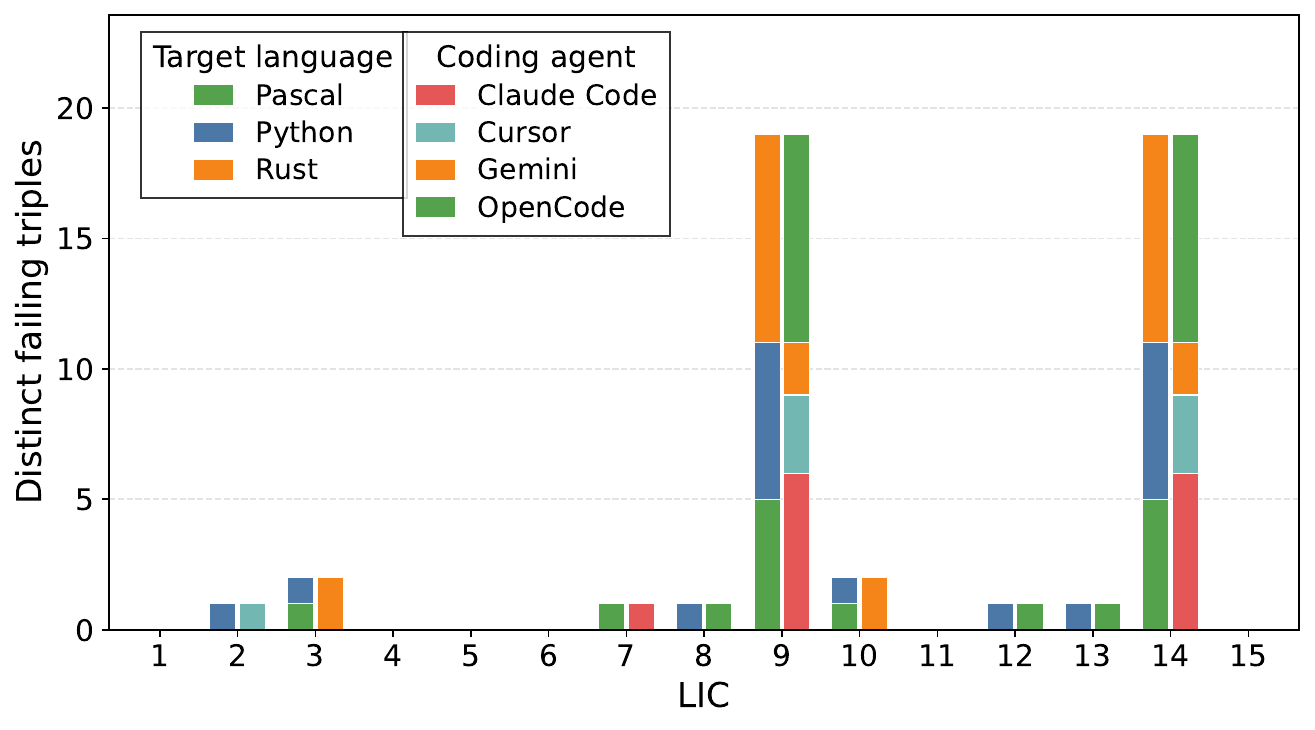}
  \caption{RQ4: failure counts per specification items (aka LIC in the LIP specification).
           For each LIC, the bars show the number of distinct
           \mbox{[harness, model, language]} triples that fail on at least one
           campaign test case. Within each LIC, the left stacked bar groups failures by target language and the right stacked bar groups the same failures by coding agent. Failures overwhelmingly concentrate in LICs 9 and 14, indicating that the failures are driven by a small number of difficult or ambiguous parts of the specification.}
  \label{fig:lic_counts}
\end{figure}

Fig.~\ref{fig:lic_counts} provides the LIC-level view.
For each LIC condition from the specification, it plots the number of distinct
\mbox{[harness, model, language]} triples that fail on at least one campaign case,
showing the same counts under two stackings: by implementation language and by coding agent.
Fig.~\ref{fig:lic_counts} shows that the failures are not equally distributed over the whole specification.

Two LICs tend to be incorrectly implemented: 9 and 14. The failures happen over all languages and across multiple agents.
Rust programs only fail on LICs \#9 and \#14, while Pascal and Python programs also fail on some other LICs.
For the admitted versions, all agents generated programs with failures, except Codex. 

LICs \#9 and \#14 are both spaced-point variants of the minimum-enclosing-circle predicate: LIC \#9 asks whether a selected triple cannot fit inside a circle of radius \texttt{RADIUS1}, while LIC \#14 strengthens that pattern into a two-part condition requiring one spaced triple outside \texttt{RADIUS1} and one spaced triple inside or on \texttt{RADIUS2}.
Inspection of the faulty versions reveals the recurring mistake behind LICs \#9 and \#14:
many implementations compute the \emph{circumcircle} of the selected triple instead of the \emph{minimum enclosing circle}.
That shortcut is incorrect because the minimum enclosing circle of three points is not always the circumcircle.
Fig.~\ref{fig:lic9-circumcircle-listing} shows an excerpt from the admitted \texttt{claude\_code/claude-sonnet-4.5} Pascal version that is representative of the mistake.

\begin{figure}[t]
\begin{lstlisting}[style=pascalcode]
function CircleContainsThreePoints(x1, y1, x2, y2, x3, y3, radius: real): boolean;
var
  d12, d23, d13: real;
  cx, cy, r: real;
  a, b, c, d, e, f, g: real;
begin
  d12 := Distance(x1, y1, x2, y2);
  d23 := Distance(x2, y2, x3, y3);
  d13 := Distance(x1, y1, x3, y3);

  if (REALCOMPARE(d12, 2 * radius) <> GT) and
     (REALCOMPARE(d23, 2 * radius) <> GT) and
     (REALCOMPARE(d13, 2 * radius) <> GT) then
  begin
    ...
    cx := (d * e - b * f) / g;
    cy := (a * f - c * e) / g;
    r := Distance(cx, cy, x1, y1);
    CircleContainsThreePoints := REALCOMPARE(r, radius) <> GT;
  end
  else
    CircleContainsThreePoints := False;
\end{lstlisting}
\caption{Representative LIC 9/14 implementation mistake: the version checks whether all pairwise distances fit within the diameter, then computes a circumcircle center and radius rather than the minimum enclosing circle.}
\label{fig:lic9-circumcircle-listing}
\end{figure}

LICs \#3 and \#10 expose a more subtle issue.
The original specification is ambiguous about whether the angle predicate should
be implemented directly as an interior angle in $[0,\pi]$ or indirectly through
a complementary angle in $[0,2\pi)$.
In exact arithmetic those formulations are equivalent,
but they are not equivalent under the benchmark's relative-tolerance
\texttt{REALCOMPARE}.
For the failing \texttt{gemini-3.1-pro-preview} versions, the oracle compares a
interior angle against the threshold $\pi-\epsilon$, while the
candidate compares the complementary angle against the threshold
$\pi+\epsilon$.
In many test cases with razor-sharp differences, comparisons against the thresholds yield different results.

Across the remaining LICs, the dominant mechanisms are concrete implementation
errors rather than specification mistakes.
Besides the minimum-enclosing-circle family in LICs \#9 and \#14, we observed a
wrong circumradius formula in LIC \#2, a segment-distance substitution for the
infinite-line distance required by LIC \#7, and dropped applicability guards in
LICs \#8, \#12, and \#13.

\begin{tcolorbox}[
  enhanced,
  colback=black!6,
  colframe=black!35,
  boxrule=0.4pt,
  arc=2pt,
  left=6pt,
  right=6pt,
  top=6pt,
  bottom=6pt
]
\noindent\textbf{Answer to RQ4.}
The dominant failure modes are concentrated in LICs \#9 and \#14, where many agents compute the circumcircle instead of the minimum enclosing circle.
The remaining recurrent faults are smaller but still structured: LICs \#3 and \#10 expose a specification ambiguity caused by \texttt{REALCOMPARE}, while LICs \#2, \#7, \#8, \#12, and \#13 fail for specific geometric or applicability-check mistakes.
These findings are consistent with Brilliant et al.'s observation that coincident failures concentrate in a small number of shared fault categories~\cite{brilliant1990analysis} and with Bishop's emphasis on specification weaknesses as drivers of correlated faults~\cite{bishop1995design}.
\end{tcolorbox}

\subsection{RQ5: N-Version Unit Reliability}
The preceding analyses show that the failures are not Bernoulli independent,
but they do not by themselves answer whether N-Version units are useful. 
We therefore construct every possible three-version unit from the admitted implementations
and evaluate each unit on the same million campaign inputs.
This yields $\binom{48}{3}=17{,}296$ triple combinations.

\begin{figure}[t]
  \centering
  \includegraphics[width=\linewidth]{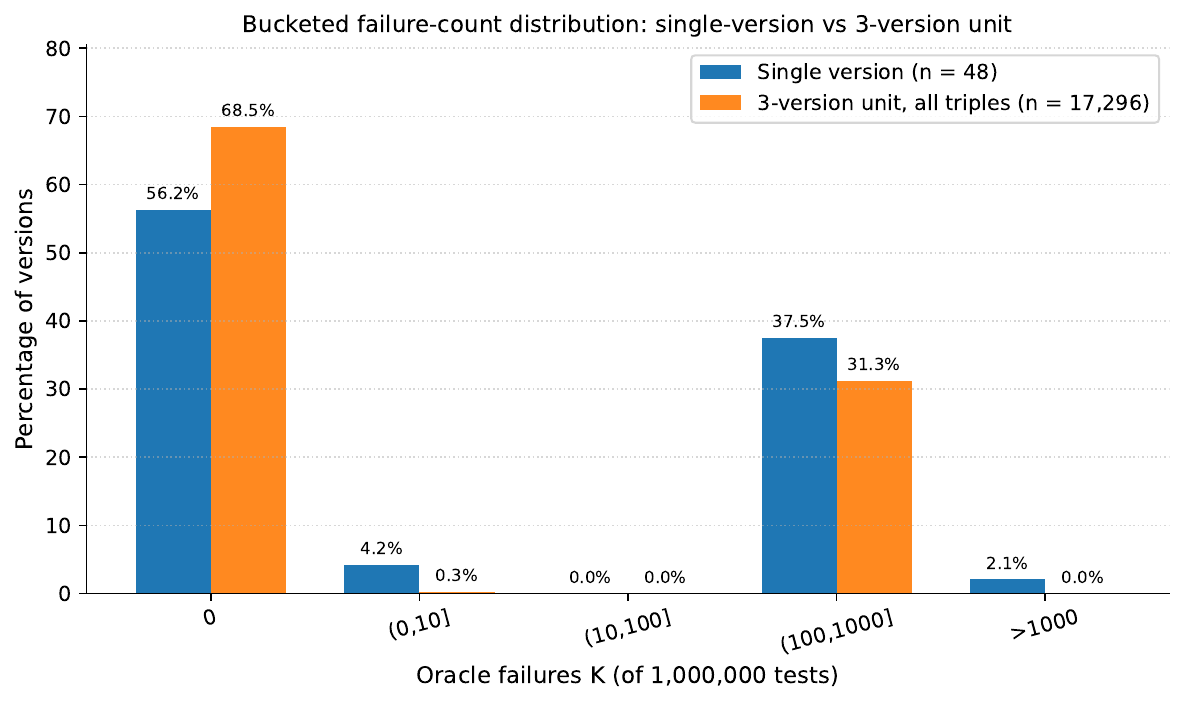}
  \caption{RQ5: Bucketed failure-count distributions for single versions and for
           all $\binom{48}{3}=17{,}296$ majority-vote triples.
           The x-axis groups versions into the buckets $0$, $(0,10]$, $(10,100]$,
           $(100,1000]$ , and $>1000$ failure counts, and the y-axis shows the
           percentage of the corresponding population in each bucket, with
           percentages annotated above the bars. The triple distribution shifts
           toward the lower-failure buckets, indicating that triple redundancy does
           reduce the number of failures even though the full
           population is not independent. This is clear evidence in favor of N-Version programming with coding agents.}
  \label{fig:all_triples_k_cdf}
\end{figure}

Fig.~\ref{fig:all_triples_k_cdf} compares bucketed failure-count distributions for
majority-vote triple failure counts against the corresponding distribution of
single-version failure counts.
The 3-version distribution is better on average: the
mean triple failure count is 130.99, compared with a mean single-version
failure count of 387.44.
At the low end, 11{,}844 triples (68.48\%) have zero majority-vote failures,
compared with 27 failure-free individual versions (56.25\%).
Both distributions have minimum 0 and median 0, so the meaningful differences appear in the upper tail: at P95 the single-version count is 429 while the triple count is 419, at P99 it is 6{,}004 versus 419, and at the maximum it is 10{,}469 versus 419.
This shows that majority voting substantially compresses the rare high-failure tail.

\begin{tcolorbox}[
  enhanced,
  colback=black!6,
  colframe=black!35,
  boxrule=0.4pt,
  arc=2pt,
  left=6pt,
  right=6pt,
  top=6pt,
  bottom=6pt
]
\noindent\textbf{Answer to RQ5.}
N-version units improve reliability even when the full population exhibit
fault correlation.
Majority-voting triples substantially improve over
single versions: the mean failure count drops from 387.44 to 130.99,
and 11{,}844 3-Version units exhibit zero observed failures (as opposed to 27 individual versions).
This is a constructive result that confirms Hatton's claim that N-Version programming is  useful, even in the presence of correlated faults~\cite{hatton1997nversion}.
\end{tcolorbox}

\section{Threats to Validity}

\textbf{Scope of Interpretation}
The present study is about \emph{implementation-level} faults relative
to a fixed specification and a fixed oracle provided by a reference implementation.
It does not test whether multi-version execution mitigate runtime or transient failures such as network outages, external tool crashes, or language-specific stack components.

\textbf{Single benchmark.}
The experiment is based on a single specification: the LIP problem.
Conclusions drawn from it may not generalize to specifications in other domains, such as long-running stateful services.
Replication over other specifications is a direction for future work.

\textbf{Sampling variability.}
Each admitted version corresponds to a single \mbox{[harness, model, language]}
tuple; we therefore measure diversity across distinct
configurations rather than within-configuration variability due to LLM sampling.
Rerunning the generation stage at either a different temperature or
at a different date would likely change the \emph{exact} failure counts.



\section{Related Work}\label{sec:related}

Recent work on LLM-based code generation has shown that diversity can improve
correctness. Chen et
al.~\cite{chen2021codex} and Li et al.~\cite{li2022alphacode} established that
sampling
multiple candidates from a single model can improve pass@k solve rates.
Kodati et al.~\cite{kodati5887028mac} further show that sampling candidates from different LLMs increases solve rates compared to single model sampling. 
In a related line of work, Wang et al.~\cite{wang2022selfconsistency} show that self-consistency
improves accuracy in code generation.
Mahmud et al.~\cite{mahmud2025ensllm} use
ensembles of LLM outputs together with syntactic and behavioral similarity
signals to improve HumanEval pass rates, and Valentin et
al.~\cite{valentin2025incoherenceoraclelessmeasureerror} argue that
cross-candidate incoherence can serve as an oracle-less proxy for error.
Differential testing of generated programs has also been suggested by Kessel et al.~\cite{kessel2024n} as a way to enhance generation quality.
These papers all exploit disagreement or agreement across candidates as useful
engineering signals. The key difference with our work is that they do not study
the reliability of a population of generated implementations under the lens of N-version programming models.

A separate thread of work emphasizes that modern software systems are often
assembled and revised under changing requirements instead of built once from a complete specification. Liu et al.~\cite{liu2026timejustintimesystemschallenges}
frame this as a move toward just-in-time systems, where specifications,
interfaces, and generated code evolve together.
This perspective is relevant to
N-version generation because repeated failures across independently generated
programs may reveal not only implementation mistakes, but also parts of the
specification that are difficult for agents to interpret
consistently.

The closest work to our setting is Galapagos~\cite{ron2025galapagos}, which consists in generating diverse but
functionally equivalent variants with guarantees.
In contrast to our work, (1) Galapagos uses a reference implementation as input to LLMs while we use a natural language specification; (2) Galapagos uses so-2023 single-shot prompting, as opposed to our usage of agentic coding.

\section{Conclusion}\label{sec:conclusion}
This paper asked whether modern AI coding agents can serve as effective
generators of diverse software versions for N-version programming.
Our results show that the answer is yes in an operational sense:
agent systems can readily generate enough working implementations to recreate fully working N-Version units.

On the question of failure independence, the resulting versions do not fail
independently, confirming \KLpair's result from TSE 1986.
Across the 48 admitted implementations in the campaign archive, the experiment
produces 429 coincident-failure cases where the independence model predicts
115.36.
Strong pairwise similarity persists across both language and agent
boundaries. That result rules out the idealized assumption or random faults  behind \KLpair's paper. Yet, this result does not make agent-generated redundancy irrelevant in practice.

First, the structure of those coincidental failures provides relevant information about the specification itself.
The implementation faults are not arbitrary: they can be repeatedly
traced back to a small set of specification weaknesses.
Second, we creating N-version units from those correlated versions, the majority voting mean failure count drops significantly.

To sum, N-Version programming with coding agents is both doable and useful. 
An interesting avenue of future research is to study how to use the correlated faults to detect and refine ambiguous specifications in an automated manner.

\bibliographystyle{ACM-Reference-Format}
\bibliography{citations}

\end{document}